\documentclass[prb,twocolumn,showpacs,floatfix,preprintnumbers,amsmath,amssymb,superscriptaddress]{revtex4}
\usepackage{graphicx}
\usepackage{dcolumn}
\usepackage{bm}
\usepackage{color}
\usepackage{color}
\usepackage[latin1]{inputenc}
\usepackage[urlcolor=blue]{hyperref}
\hypersetup{backref, colorlinks=true} 
\begin{document}

\preprint{APS/123-QED}

\title{Lower critical field and SNS-Andreev spectroscopy of 122-arsenides: Evidence of nodeless superconducting gap}

\author{M. Abdel-Hafiez}\email{m.mohamed@hpstar.ac.cn}
\affiliation{Center for High Pressure Science and Technology Advanced Research, 1690 Cailun Rd., Shanghai, 201203, China}
\affiliation{D\'epartement de Physique, Universit\'e de Li\`ege, B-4000 Sart Tilman, Belgium}
\affiliation{Faculty of science, Physics department, Fayoum University, 63514-Fayoum- Egypt}

\author{P. J. Pereira}\affiliation{INPAC, Catholic University of Leuven, Celestijnenlaan 200D, B--3001 Leuven, Belgium}
\author{S. A. Kuzmichev}
\affiliation{Low Temperature Physics and Superconductivity Department, Physics Faculty, M.V. Lomonosov Moscow State University, 119991 Moscow, Russia}
\author{T. E. Kuzmicheva}
\affiliation{P. N. Lebedev Physical Institute, Russian Academy of Sciences, Moscow 119991, Russia}
\author{V. M. Pudalov}
\affiliation{P. N. Lebedev Physical Institute, Russian Academy of Sciences, Moscow 119991, Russia}
\affiliation{Moscow Institute  of Physics and Technology, Moscow 141700, Russia}
\author{L. Harnagea}\affiliation{Leibniz-Institute for Solid State and Materials Research, (IFW)-Dresden, D-01171 Dresden, Germany}
\author{A. A. Kordyuk}
\affiliation{Institute of Metal Physics of National Academy of Sciences of Ukraine, 03142 Kyiv, Ukraine}
\author{A. V. Silhanek}
\affiliation{D\'epartement de Physique, Universit\'e de Li\`ege, B-4000 Sart Tilman, Belgium}
\author{V. V. Moshchalkov}
\affiliation{INPAC, Catholic University of Leuven, Celestijnenlaan 200D, B--3001 Leuven, Belgium}

\author{B. Shen}
\affiliation{Institute of Physics, Chinese Academy of Sciences, Beijing 100190, China}
\author{Hai-Hu Wen}
\affiliation{National Laboratory for Solid State Microstructures and Department of Physics, Nanjing University, Nanjing 210093, China}

\author{A. N. Vasiliev}
\affiliation{Low Temperature Physics and Superconductivity Department, Physics Faculty, M.V. Lomonosov Moscow State University, 119991 Moscow, Russia}
\affiliation{Theoretical Physics and Applied Mathematics Department, Ural Federal University, 620002 Ekaterinburg, Russia}
\author{Xiao-Jia Chen}
\affiliation{Center for High Pressure Science and Technology Advanced Research, 1690 Cailun Rd., Shanghai, 201203, China}
\date{\today}

\begin{abstract}
Using two different experimental techniques, we studied  single crystals of  the 122-FeAs family with almost the same critical temperature, $T_{c}$. We investigated the temperature dependence of the lower critical field $H_{c1}(T)$ of a Ca$_{0.32}$Na$_{0.68}$Fe$_{2}$As$_{2}$ ($T_{c} \approx$ 34 K) single crystal under static magnetic fields $H$ parallel to the $c$ axis. The temperature dependence of the London penetration depth can be described equally well either by a single anisotropic $s$-wave-like gap or by a two-gaps model, while a {$d$-wave approach cannot be used to fit the London penetration depth data. The intrinsic multiple Andreev reflection effect (IMARE) spectroscopy was used to detect bulk gap values in single crystals {of the intimate compound} Ba$_{0.65}$K$_{0.35}$Fe$_{2}$As$_{2}$, with the same $T_{c}$. We estimate the range of the large gap value $\Delta _{L}$ = (6--8) meV (depending on small variation of $T_{c}$) and its $k$-space anisotropy about 30\%, and the small gap is about $\Delta _{S}\approx$ (1.7$\pm$0.3) meV. {This {clearly} indicates that the gap structure of our investigated systems is more likely to be of the two-gap nodeless $s$-wave type.}}
\end{abstract}


\maketitle

\section{Introduction}
Fe-based superconductors of the $A$Fe$_{2}$As$_{2}$ type (122 system), where $A$ is an alkaline-earth element (i.e., Ca, Ba, Sr) show an intermediate critical temperature $T_{c}$, high upper critical fields $H_{c2}$ due to the small coherence lengths and low anisotropy ($\gamma \approx$ 2).~\cite{pag} The identification of the symmetry and structure of the superconducting order parameter and the mechanism for Cooper pairing is of primary importance in Fe-based superconductors. Numerous efforts were made since the discovery of high-$T_{c}$ Fe-based superconductors in order to understand the physics of the pairing mechanism. It turns out that the physics of the pairing could be more complicated than it was originally thought, because of the multi-band nature of low-energy electronic excitations.~\cite{Ch} In quasi-2D multiband superconductors two or more energy bands at the Fermi energy give rise to multiple energy gaps in the respective superconducting condensates.~\cite{pag,HJC,VAM,HSU} Recent specific heat and angle-resolved photoemission spectroscopy (ARPES) measurements provide clear evidence of multiple gap structures in 122 system.~\cite{DE,STJ}

On the other hand, the density of states (DOS) calculations show that the states at the Fermi level E$_{F}$ are formed mainly by 3$d$-electrons of Fe, thus the metallic-type conductance is namely due to these 3$d$-states.~\cite{DJS,DK} This leads to the suggestion that any kind of spacer between FeAs blocks affects the level of doping rather than fundamental pairing mechanism. Consequently, one could assume that spacer doping has a minor influence on the superconducting gap symmetry.

Various experimental data on gap magnitude and anisotropy in $k$-space are contradictory enough. In particular, the specific heat (SH) measurements are commonly used for  gap quantifying,~\cite{STJ,AK,Popovich2010} though there are several known problems with data treatment. The SH data contains contribution from the lattice, that is subtracted to some extent in order to determine the electronic contribution. The lattice contribution to the SH is typically estimated by suppressing the superconducting transition in high magnetic fields. Therefore, the lattice SH cannot be accurately obtained because of the very high upper critical field of the hole-doped and magnetic/structural phase transitions at higher temperature of the parent compound. The majority of the earlier SH data suffer from a residual low-temperature non-superconducting electronic contribution and show Schottky anomalies.~\cite{AK,HD} Moreover, superconductivity-induced electronic SH is very sensitive to the sample quality and phase purity.~\cite{Popovich2010} Also, in the earlier SH data analysis, the data are commonly fitted  to the  phenomenological multiband model,~\cite{Pad} that assumes a BCS temperature dependence of the gaps. However, our Andreev spectroscopy measurements~\cite{SNT,SNT2,SAK,TEK2,MGM} does not support this assumption and clearly show that the $\Delta(T)$ dependencies for the multiband superconductors (such as MgB$_{2}$, and Fe-based superconductors) deviate substantially from the BCS-type because of the interband coupling. Finally,  fitting the SH data with the multiband model requires several adjustable parameters. It might be therefore that a combination of all the above obstacles causes  dissimilar gap values (and, in particular, their unrealistically large values, such as 11 and 3.5 meV~\cite{Popovich2010}), obtained from the SH measurements. In this context, of highly importance is to have the possibility of comparing  the results obtained by two independent bulk purely electronic probes; particular good candidates are the the lower critical field, $H_{c1}$, and Andreev spectroscopy.


The determination of $H_{c1}$, the field at which vortices penetrate into the sample, allows one to extract the magnetic penetration depth, $\lambda$, a fundamental parameter characterizing the superconducting condensate and carrying information about the underlying pairing mechanism. In the superconducting state, the temperature dependence of the penetration depth is a sensitive measure of low-energy quasiparticles, making it a powerful tool for probing the superconducting gap.~\cite{VGK} The lower critical field studies in LiFeAs,~\cite{KSS,YS} Ba$_{0.6}$K$_{0.4}$Fe$_{2}$As$_{2}$,~\cite{CR} Eu$_{0.5}$K$_{0.5}$Fe$_{2}$As$_{2}$,~\cite{An} and FeSe~\cite{Hafiez1} have supported the existence of two $s$-wave-like gaps. On the other hand, nodes in the SC gap have been reported in NdFeAsO$_{0.82}$F$_{0.18}$ and La-1111, where the magnetic penetration depth exhibited a nearly linear temperature dependence.~\cite{XLW,Hicks} Also, a nodal pairing state of Sm-1111 has been suggested based on the $T^{2}$ dependence of the $H_{c1}$ studies.~\cite{Sm} The interpretation of these results may also be impaired by substantial contributions from paramagnetic centers. In the view of existing divergency  of conclusions about the gap symmetry derived from single-type measurements,  there is a clear need of obtaining the set of data by different techniques.

Although the superconducting order parameter has been investigated for similar compounds, i.e., Ba$_{0.65}$Na$_{0.35}$Fe$_{2}$As$_{2}$~\cite{AK} and Ba$_{0.68}$K$_{0.32}$Fe$_{2}$As$_{2}$,~\cite{Popovich2010} its investigations on the Ca$_{0.32}$Na$_{0.68}$Fe$_{2}$As$_{2}$ and Ba$_{0.65}$K$_{0.35}$Fe$_{2}$As$_{2}$  are necessary in order to further clarify the differences between these structurally similar systems. It has been recently shown in Ref.~\onlinecite{DE} that Ca$_{0.32}$Na$_{0.68}$Fe$_{2}$As$_{2}$ is almost identical to the more studied Ba$_{0.65}$K$_{0.35}$Fe$_{2}$As$_{2}$ in terms of electronic band structure, Fermi surface topology, and superconducting gap distribution. Therefore we were able to combine here the results for these two compounds measured by different techniques. In this study we investigated whether both London penetration depth and intrinsic multiple Andreev reflection effect (IMARE) techniques may provide such conclusive and self-consistent information on the gap anisotropy. Based on our experimental data, we report on the superconducting gap properties of the hole-doped Ca$_{0.32}$Na$_{0.68}$Fe$_{2}$As$_{2}$ and Ba$_{0.65}$K$_{0.35}$Fe$_{2}$As$_{2}$. Our analysis shows that the superconducting gaps determined through fitting to the London penetration depth for the out-of-plane directions support two possible scenarios, namely, the presence of anisotropic single gap and the two $s$-wave-like gaps with different magnitudes and contributions. In addition, our IMARE spectroscopy of SNS-Andreev arrays formed by the break-junction technique for Ba$_{0.65}$K$_{0.35}$Fe$_{2}$As$_{2}$ reveals two nodeless gaps: the large gap,  $\Delta _{L}$ = (6--8) meV with extended $s$-wave symmetry in the $k$-space, and the small gap, $\Delta _{S}$ = 1.7$\pm$0.3 meV.


\section{Experimental details}

The DC magnetization measurements discussed in this paper were performed on a rectangular slab. The single crystals of Ca$_{0.32}$Na$_{0.68}$Fe$_{2}$As$_{2}$ were grown using NaAs as described in Ref.~\onlinecite{LH}. The chemical composition was verified by scanning electron microscope (SEM-Philips XL 30) equipped with an energy dispersive X-ray (EDX) spectroscopy probe. The magnetization measurements were performed by using a superconducting quantum interference device magnetometer (MPMS-XL5) from Quantum Design. The good quality of the crystals was confirmed from various physical characterizations: (i) A sharp specific heat anomaly associated with the superconducting phase transition is observed at 34\,K~\cite{STJ} (ii) The large value of the residual resistivity ratio (RRR) is found to be $\rho_{(300\,K)}$/$\rho_{(36\,K)}$ = 12.8 confirming the good quality of the single crystal~\cite{STJ} (iii) This system also stands out due to the absence of nesting between hole and electron pockets of the Fermi surface.~\cite{DE} Ba$_{0.65}$K$_{0.35}$Fe$_{2}$As$_{2}$ single crystals were synthesized by self-flux method using FeAs as the flux, for details see.~\cite{Shan,HQL} The chemical composition and crystal structure were checked by X-ray diffraction and energy dispersive X-ray microanalysis. For both compounds a critical temperature $T_{c} \approx$ 34\,K,  is evidenced by magnetization measurements for Ca$_{0.32}$Na$_{0.68}$Fe$_{2}$As$_{2}$, and by Andreev spectra flatting in case of Ba$_{0.65}$K$_{0.35}$Fe$_{2}$As$_{2}$ study.

The single crystal prepared for Andreev spectroscopy studies is a thin plate of about $a \times$$b \times$$c$ = (2--4)$\times$(1--2)$\times$(0.05--0.15)\,mm$^{3}$. The crystal was attached to a spring sample holder by four liquid In-Ga pads (true 4-contact connection), thus making ab-plane to be parallel to the sample holder, and then cooled down to $T$ = 4.2 K. Next, the sample holder was curved mechanically ("break-junction" technique~\cite{JMO}). Under such deformation, the single crystal was cracked generating the superconductor - constriction (weak link) - superconductor contact (ScS). Since the microcrack was located deep in the bulk of the sample and remotely from current leads, the cryogenic clefts were free of overheating and degradation caused by impurity penetration if any. Multiple Andreev reflection effect (MARE) is observed in ballistic constrictions of the metallic type, where the diameter $2a$ of the area is less than the quasiparticle mean free path $l$.~\cite{YuV,AFA} MARE manifests itself causing an excess current at low biases in current-voltage characteristic (CVC) of ScS contact. With it a series of dynamic conductance peculiarities called subharmonic gap structure (SGS) appears. The position V$_{n}$ of such peculiarities is determined by the superconducting gap, V$_{n}$ = 2$\Delta$/$en$ (n-natural number)~\cite{GBA,Flen} at any temperatures up to $T_{c}$.~\cite{MOc,AKU} For the high-transparency SnS-Andreev regime typical for our break-junction contacts, the supercurrent is absent, whereas the SGS represents dynamic conductance dips for the gap of both $s$- and $d$-wave symmetries.~\cite{AKU,AP} The coexistence of two superconducting gaps would cause, obviously, two SGS's in the dI/dV-spectrum.

Now we estimate the typical diameter of the contacts formed in Ba-122 single crystals under study. The product of the normal-state bulk resistivity $\rho_n$ and quasiparticle mean free path $l$ was shown to be from $\rho_n l \approx 0.45 \times 10^{-9}$ \,Ohm$\cdot$cm$^2$~\cite{VNZ} to $\sim 1.7 \times 10^{9}$\,Ohm$\cdot$cm$^2$,~\cite{Machida} determining the range of the experimental uncertainty. Taking the value of the in-plane bulk resistivity $\rho_n^{ab} \approx 0.4 \times 10^{-5}$\,Ohm$\cdot$cm for our Ba-122 single crystal (as described in,~\cite{BSHY,ZSW} and the anisotropy value $\gamma^{H_{c2}} = \sqrt{\gamma^{\rho}} \approx 1.8$,~\cite{ZSW} we determine the $c-$axis resistivity as $\rho_n^c \sim 1.3 \times 10^{-4}$\,Ohm$\cdot$cm. Hence, for the $\rho_n l \approx 0.45 \times 10^{-9}$\,Ohm$\cdot$cm$^2$~\cite{VNZ} and for the Ba-122 samples used the quasiparticles mean free path along the c-direction is $l^c \approx 35$\,nm. In case of $\rho_n l \approx 1.7 \times 10^{-9}$ \,Ohm$\cdot$cm$^2$ [*ref. 40] one gets $l^c \approx 133$\,nm. Finally, using Sharvin formula for ballistic contact and the typical resistance of our SnS-contact, $R = 10 \div 100$\,Ohm, we calculate the contact radius as $a = \sqrt{\frac{4}{3\pi} \frac{\rho_nl}{R}} = 13 \div 82$\,nm. This rough estimation shows moderate superiority of $l^c$ over $a$, thus proving the ballistic regime and allowing observation of $1 \textendash 3$ Andreev peculiarities in $dI/dV$-spectra.~\cite{AKU}


Due to its layered structure, Ba-122 single crystal cleaves along ab-planes with steps-and-terraces on cryogenic clefts. Each step is a natural stack of superconducting Fe-As blocks separated with metallic Ba spacers, and in fact, represents an S-n-S-n-$_{...}$-S array. In ballistic mode, these arrays manifest intrinsic multiple Andreev reflections effect (IMARE)~\cite{Pono} which is similar to the intrinsic Josephson effect in cuprates.~\cite{BAA} In our experiment, after cleavage the two superconducting banks of the sample slide over each other touching at different terraces. By precise tuning of the sample holder one can probe tens of SnS-contacts as well as arrays (containing various number of contacts) in order to check reproducibility of the gap values. Since the S-n-S-n-$_{...}$-S stack contains a sequence of $N$ connected junctions (with the transport along the $c$-direction), the SGS dips appear at bias voltages being $N$ times larger:
 {\begin{equation}
 V_{n} = \frac{2\Delta _{i}N}{en},
\end{equation}}
so do other peculiarities caused by $bulk$ properties of material. The number of junctions $N$ could be determined by normalization the spectrum of array contact to that of single SnS-contact; then the positions of each gap SGS should coincide. Probing such natural arrays, one obtains information about the true bulk properties of the sample (almost non-affected by surface states which seem to be significant in Ba-122~\cite{EvH}) $locally$ (within the contact  size $a \approx$ 20--80 nm). This feature favors accuracy increasing in gap magnitude measurements.~\cite{TEK3}

To get dI(V)/dV spectra directly, we use current generator that mixes DC with the low-amplitude AC. In this way the standard modulation method is used. Narrowband signal amplification with the help of Lock-in nanovoltmeter reveals dI/dV peculiarities even in case of the nearly flat I(V) characteristics.


\section{Results and discussions}

\subsection{Irreversible magnetization}

\begin{figure}
\includegraphics[width=20pc,clip]{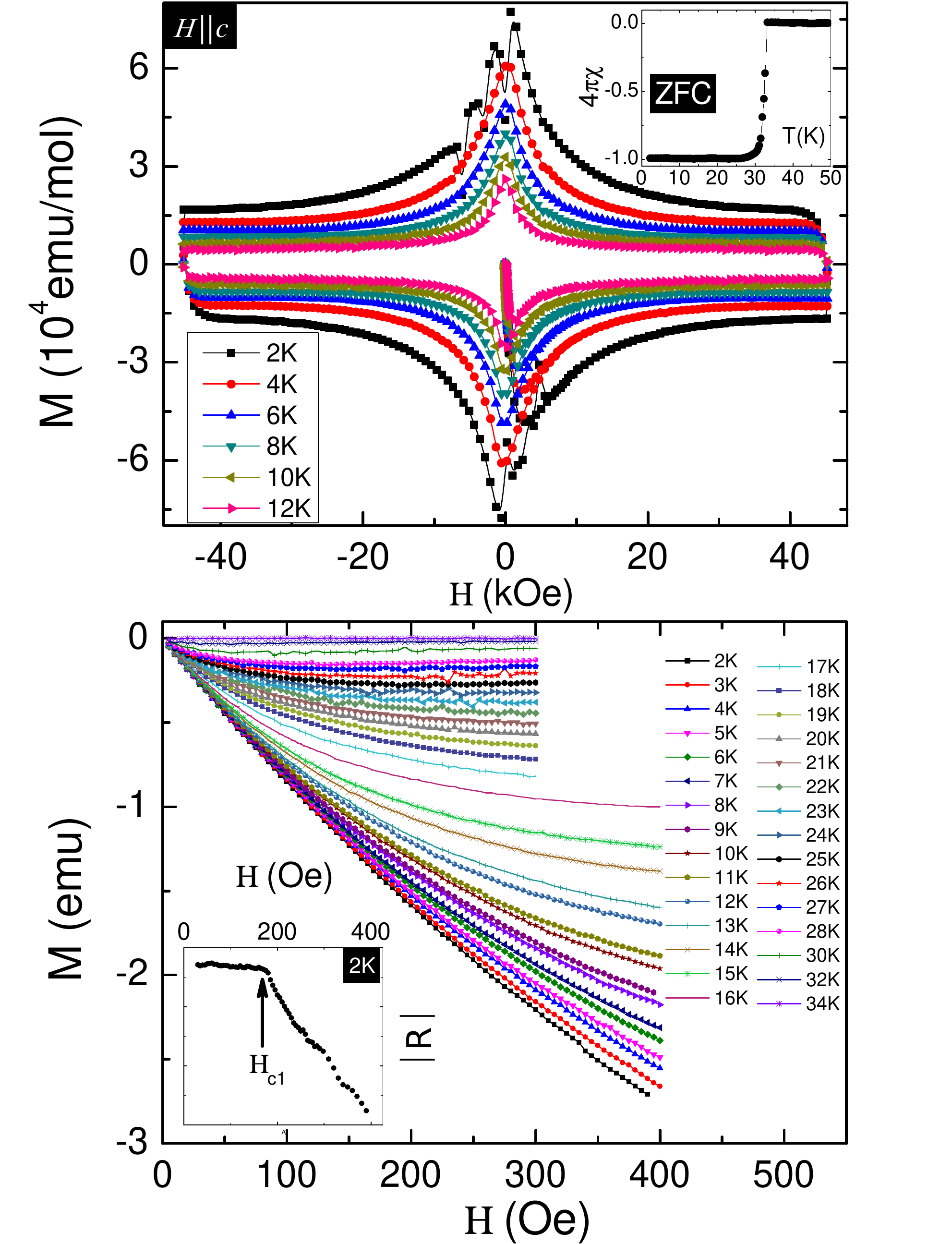}
\caption{(Upper panel) Magnetic field dependence of the isothermal magnetization $M$ vs. $H$ loops measured at different temperatures ranging from 2 to 12\,K up to 45\,kOe with the field parallel to the $c$ axis for Ca$_{0.32}$Na$_{0.68}$Fe$_{2}$As$_{2}$ single crystal. The inset shows the temperature dependence of the ZFC magnetic susceptibility $\chi$ after demagnetization correction in an external field of 10\,Oe applied along $c$. (lower panel) The initial part of the magnetization curves measured at various temperatures for $H \parallel c$. The inset depicts an example used to determine the $H _{c1}$ value using the regression factor, $R$, at $T$ = 2\,K} \label{Fig:1}
\end{figure}

Figure\,1(upper panel) presents the field dependence of the isothermal magnetization $M$ at various temperatures up to 45\,kOe for $H \parallel c$. At $T$ = 2\,K,  the $M(H)$ exhibits irregular jumps close to $H$ = 0  {similarly to LiFeAs  and Ba$_{0.65}$Na$_{0.35}$Fe$_2$As$_2$ superconductors.~\cite{Ashim,Ashim1}} These flux jumps are usually attributed to thermoelectromagnetic instability.~\cite{MEM} The inset of Fig.\,1 (upper panel) shows the temperature dependence of the magnetic susceptibility ($\chi = M/H$) measured by following zero-field cooled (ZFC) procedures in an external field of 10\,Oe applied along $c$ axis. The DC magnetic susceptibility exhibits a superconducting temperature transition with an onset at 34\,K. It is worth mentioning that our system exhibits a strong bulk pinning reflected by the symmetric hysteresis loops about the horizontal axis $M=0$. In addition, the superconducting $M(H)$ exhibits no magnetic background. This indicates that the sample contains negligible magnetic impurities. The virgin $M(H)$ curves at low fields at several temperatures are collected in Fig.\,1 (lower panel) for $H \parallel c$. {In order to determine the transition from linear to non-linear $M(H)$, a user-independent procedure consisting of calculating the regression coefficient $R$ of a linear fit to the data points collected between $0$ and $H$, as a function of $H$ is used. Then, $H_{c1}$ is taken as the point where the function $R(H)$ starts to deviate from linear dependence. This procedure is similar to that previously used in the studies shown in the Ref.~\onlinecite{Hafiez1} and illustrated for a particular temperature $T = 2$\,K in the inset of the lower panel of Fig.\,1.} From the magnetization hysteresis loops $M(H)$, we calculated the critical current density $J_{c}$ by using the critical state model with the assumption of field-independent $J_{c}$. We obtain $J_{c} \sim$ 1.15 $\cdot$10$^{6}$ A/cm$^{2}$ for $H \parallel c$ at 2\,K, see Fig.\,2. The inset of Fig.\,2 demonstrates a strong temperature dependence of $J_{c}(H=0)$. In addition, the error bar at $T$ =2\,K shows the uncertainty of the estimated value due to the irregular jumps close to $H$ = 0.

\begin{figure}
\includegraphics[width=20pc,clip]{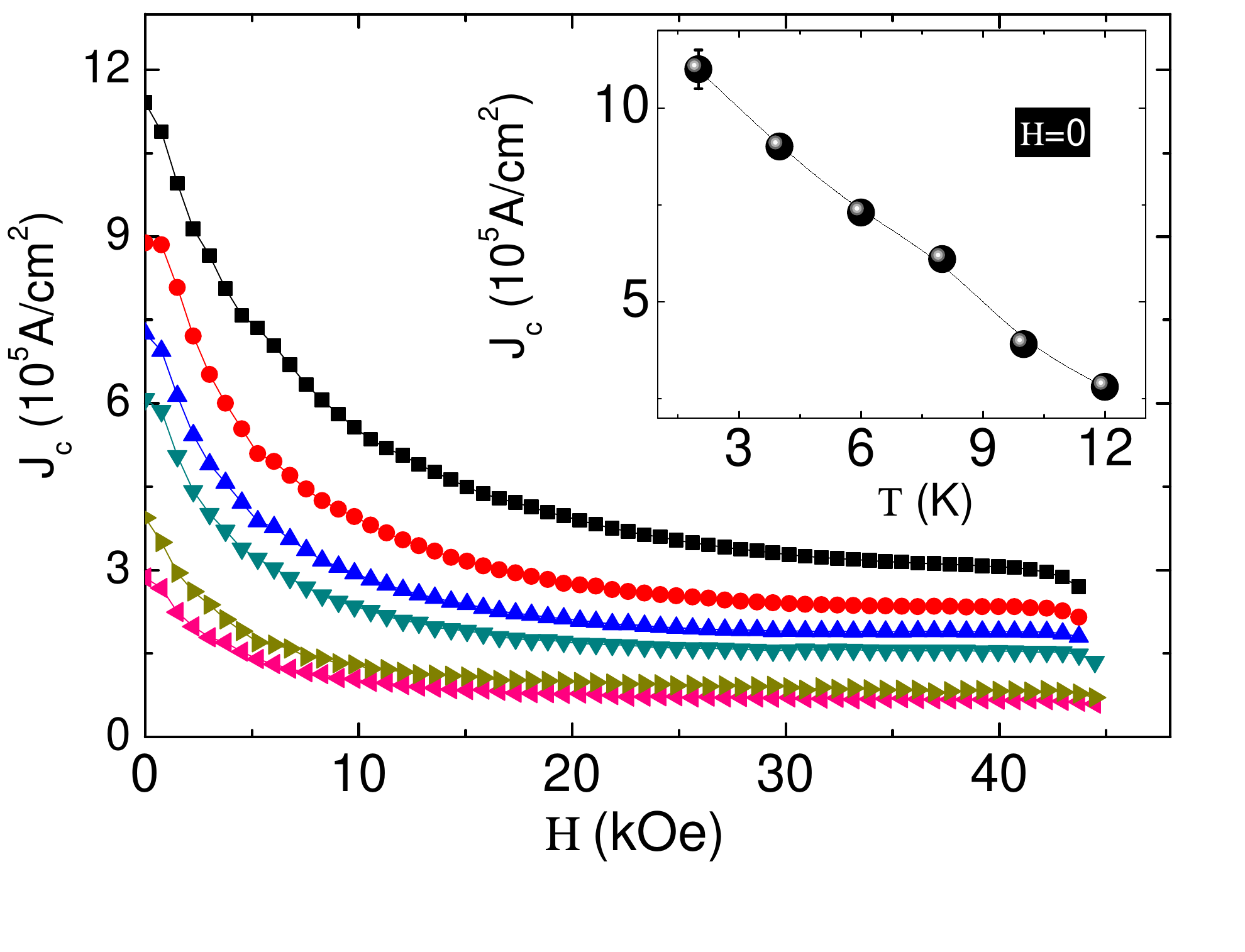}
\caption{The critical current density $J_{c}$ at various temperatures up to 45\,kOe for $H \parallel c$. The inset presents the temperature dependence of the $J_{c}$ values at $H$ = 0 for the Ca$_{0.32}$Na$_{0.68}$Fe$_{2}$As$_{2}$ single crystal. The line is a guide to the eyes. The error bar at $T$ =2\,K shows the uncertainty of the estimated value due to the irregular jumps close to $H$ = 0.} \label{Fig:1}
\end{figure}

Due to the high sensitivity, the measurements on bulk single crystals will detect first flux lines penetration into areas with large demagnetizing fields such as sharp corners, edges or inclusions of the normal state defects.~\cite{Moshchalkov} In addition, determining $H_{c1}$ from magnetization measurements is not always reliable, since this type of effect can mask completely the predicted sharp drop in the magnetization at $H_{c1}$. A popular approach to measure  $H_{c1}$ consists of measuring the magnetization $M$ as a function of $H$ and identifying the deviation of the linear Meissner response which would correspond to the vortex penetration. This technique implicitly relies on the assumption that no surface barriers are present, thus assuring that $H_{c1}$ coincides with vortex penetration field. The $H _{c1}$ values illustrated in the main panel of Fig.\,3 for $H \parallel c$ show the most intriguing feature which is the upward trend with negative curvature over the entire temperature range 0-$T_{c}$. Similar trend is reported for Ba$_{0.6}$K$_{0.4}$Fe$_{2}$As$_{2}$~\cite{CR} and FeTe$_{0.6}$Se$_{0.4}$.~\cite{Yadav} The inset of Fig.\,3 shows the normalized temperature dependence, $H_{c1}(T)/H_{c1}(0)$ versus $T/T_{c}$, of Ca$_{0.32}$Na$_{0.68}$Fe$_{2}$As$_{2}$ together with various systems of Fe-based superconductors,~\cite{CR,KSS,YS,Yadav,Hafiez1,Nd,Pr,Sm} MgB$_{2}$,~\cite{Mg} and YBa$_{2}$Cu$_{3}$O$_{6+x}$.~\cite{Cu}

In the London theory, the penetration depth, $\lambda(T)$ = $\lambda(T=0)$+$\delta\lambda(T)$ behaves as $\delta\lambda(T) \propto\exp(\frac{-\Delta}{\kappa _{B}T})$ at low $T$ in the $s$-wave pairing with a true gap everywhere on the Fermi surface, reflecting superconducting gap $\Delta$. In $d$- wave pairing with line nodes, $\delta\lambda(T) \propto T$ at low $T$ in the clean limit. These indicate that $H_{c1}(T)$ depends on the pairing symmetry of anisotropic superconductors. In order to shed light on the pairing symmetry in our system, we estimated the penetration depth at low temperatures using the traditional Ginzburg-Landau (GL) theory, where $H_{c1}$ is given by:~\cite{ALF} $\mu_{0}H_{c1}^{\parallel c} = (\phi$$_{0}/4\pi\lambda _{ab}^{2})\ln\kappa _{c}$, where $\phi$$_{0}$ is the magnetic-flux quantum $\phi$$_{0}$ = $h/e^{\ast}$ = 2.07 x 10$^{-7}$Oe cm$^{2}$, $\kappa _{c}$ =$\lambda _{ab}$/$\xi _{ab}$ is the Ginzburg-Landau parameter. The value of $\kappa$ was determined from the equation:$\frac{2H_{c1}(0)}{H_{c2}(0)} = \frac{\ln\kappa+0.5}{\kappa^{2}}$. Solving this equation numerically for Ca$_{0.32}$Na$_{0.68}$Fe$_{2}$As$_{2}$ using the values $H_{c1}(0)$ and  {$H_{c2}(0)$, which is taken from specific heat data as reported in Ref.~\cite{STJ}}, we obtained $\kappa _{c}$ = 139. Using this value of $\kappa$, we obtained $\lambda$(0) = 212 (10) nm. This value is in close agreement with the values reported for Ba(Fe$_{0.93}$C$_{0.07}$)$_{2}$As$_{2}$ ($\lambda$(0) = 208 nm),~\cite{RTG} LiFeAs ($\lambda$(0) = 198.4 nm),~\cite{YS} La-1111 ($\lambda$(0) = 245 nm),~\cite{Lutk} and Sm-1111 ($\lambda$(0) = 190 nm).~\cite{AJD}

\begin{figure}
\includegraphics[width=19pc,clip]{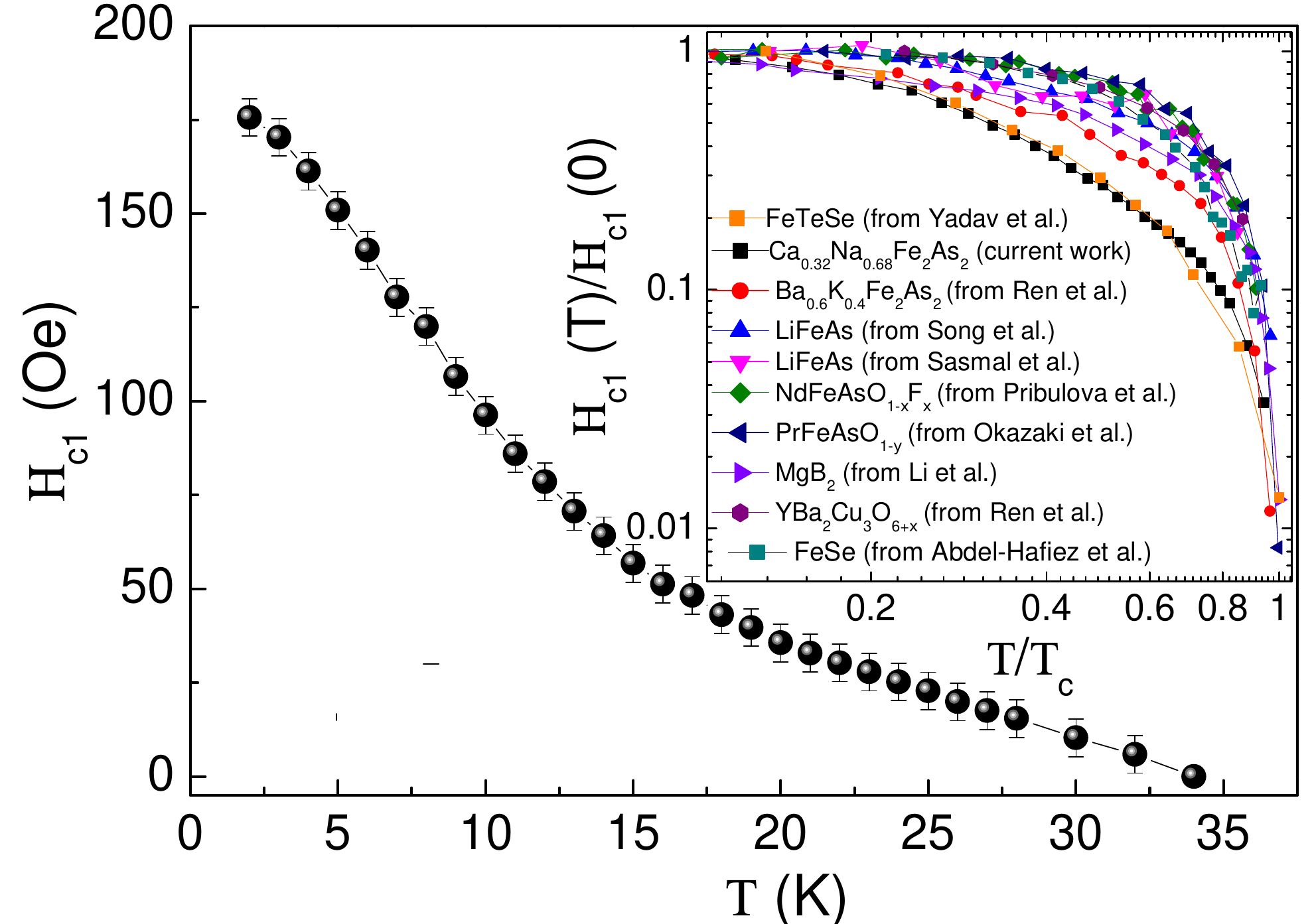}
\caption{The main panel shows the temperature dependence of $H_{\mathrm{c1}}$ vs. temperature for the field applied parallel to $c$ axis. $H_\mathrm{c1}$ has been estimated from the regression factor (see the inset of the lower panel in Fig.\,2). The bars show the uncertainty of estimated by the deviating point of the regression fits. The inset shows the scaling of the lower critical field values of Ca$_{0.32}$Na$_{0.68}$Fe$_{2}$As$_{2}$ together with various Fe-based, MgB$_{2}$, and cuprates superconductors (see text).} \label{Fig:1}
\end{figure}
\subsection{Theoretical fitting of the lower critical field}

\begin{figure}
\includegraphics[width=19pc,clip]{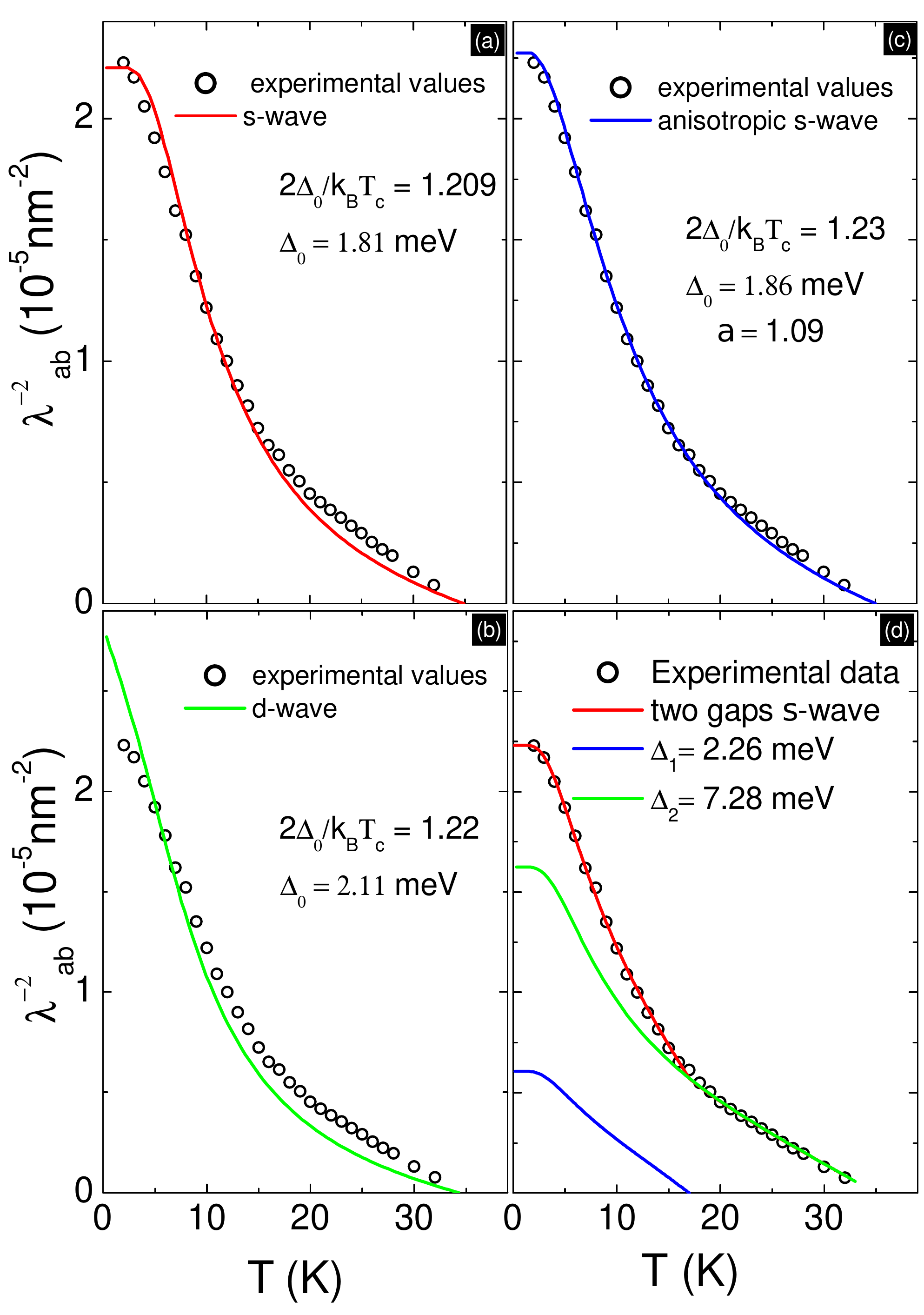}
\caption{Plots showing the fitting of theoretical curves (described by the method in the text) to experimental data of the temperature dependence of $\lambda^{-2}_{ab}$ calculated using $H_{c1}$ for the field applied parallel to $c$ axis. For single $s$-wave, BCS fit, $d$-wave, anisotropic $s$-wave and two-gap $s$-wave, the fit through the experimental data is shown as solid lines in (a), (b), (c), and (d) panels, respectively. Also, in the (d) panel, it can be seen two solid lines, that represent the contribution by the the gap $\Delta_1(0)$ and $\Delta_2(0)$ to fitting of $\lambda^{-2}_{ab}$.} \label{Fig:1}
\end{figure}

Up to date, concerning the pairing symmetry in Fe-based superconductors, the debate is wide open and various scenarios are still under discussions. For instance, different experimental results are divided between those supporting line nodes~\cite{Hafiez2,JS} and isotropic as well as anisotropic nodeless gaps,~\cite{NXu,KH,Miao,QQ,Allan,TK,Yapono,SAK,TEK2,TEK3} and two-gap superconductivity.~\cite{AK,Popovich2010,KH2} Taking this into account, the obtained experimental temperature dependence of $\lambda^{-2}(T)$ were analyzed by using the phenomenological $\alpha$-model, see Fig.\,4. This model generalizes the temperature dependence of gap to allow $\alpha=2 \Delta(0)/T_c > 3.53$ (i.e. $\alpha$ values higher than the BCS value), taking into account the behavior of this function in strong coupling regime. The temperature dependence of each energy gap for this model can be approximated as:~\cite{Carrington} $\Delta _{i}(T) = \Delta _{i}(0) {\tanh[1.82(1.018(\frac{T_{ci}}{T}-1))^{0.51}]}$, where $\Delta(0)$ is the maximum gap value at $T$ = 0. We fit the temperature dependence of the London penetration depth using the following expression:
\begin{equation}
 \frac{\lambda _{ab}^{-2}(T)}{\lambda _{ab}^{-2}(0)} = 1+\frac{1}{\pi}\int^{2 \pi}_0{ 2\int_{\Delta(T,\phi)}^{\infty}{\frac{\partial f}{\partial E} \frac{E dE d\phi}{\sqrt{E^2-\Delta^2(T,\phi)}}}},
\end{equation}
where $f$ is the Fermi function $ [ \exp( \beta E + 1 )]^{-1}$, $\varphi$ is the angle along the Fermi surface, $\beta$ =
($k_\textup{B}T)^{-1}$. The energy of the quasiparticles is given by $E$ = $[\epsilon^{2} + \Delta^{2}(t)]^{0.5}$, with $\epsilon$ being the energy of the normal electrons relative to the Fermi {level}, and where $\Delta(T,\phi)$ is the order parameter as function of temperature and angle. For different types of order parameter symmetries (e.g., $d$-wave, anisotropic $s$-wave, etc...) we have different angular dependencies of the order parameter. Thus, the experimental data were analyzed by using $s$-wave, $d$-wave, anisotropic $s$-wave the following expressions  $\Delta(T,\phi)=\Delta(T)$, $\Delta(T,\phi)=\Delta(T) \cos(2 \theta)$, and $\Delta(T,\phi) = (1+a\cos(\theta))/(1+a)$ respectively, where $a$ is the anisotropy parameter. For the two-gap model, $\lambda^{-2}_{ab}$ is calculated as:~\cite{Carrington}
 {\begin{equation}
\lambda _{ab}^{-2}(T) = r\lambda _{1}^{-2}(T) + (1-r)\lambda _{2}^{-2}(T),
\end{equation}}
 where $0<r<1$.

{The best description of the experimental data for each type of order parameter, single gap $s$-wave, $d$-wave, anisotropic $s$-wave and two-gaps $s$-wave can be seen in Fig.\,4(a), (b), (c), and (d) respectively. The corresponding gap values are shown inside of these plots. The main features in Fig.\,4 can be described in the following way: (i) As a first step we compare our data to the single band $s$-wave and we find a systematic deviation at high temperature data, see Fig.\,4(a) (ii) More obvious deviations exist in the case of $d$-wave approach as shown in Fig.\,4(b). This {clearly} indicates that the gap structure of our system is more likely to be nodeless $s$-wave, which compares reasonably well with pervious experimental ARPES data.~\cite{DE}} (iii) Then, both anisotropic $s$-wave and the two-gaps model are further introduced to fit the experimental data. For the anisotropic $s$-wave, the fitting  with the magnitude of the gap $\Delta _0$ = 1.86\,meV is shown in Fig.\,4(c) with an anisotropy parameter $\approx$1.09. As can be seen the anisotropic $s$-wave order parameter presents a well description to the data. (iv) Equally good description of the experimental data for the two-gaps $s$-wave model is obtained using values of $\Delta _{1}(0)$ = 2.26\,meV, $\Delta _{2}(0)$ = 7.28\,meV. Equations (1) and (2) are used to introduce the two gaps and their appropriate weights. However, we remind the reader that in this approach the one-band expression is generalized to the two-band case. The gap values for each gap are shown individually in Fig.\,4(d). It is noteworthy that our extracted gap values are comparable with the two-band $s$-wave fit, $\Delta _{1,2}(0)$= 2.2 and 8.8\,meV, reported for Ba$_{0.6}$Ka$_{0.4}$Fe$_{2}$As$_{2}$.~\cite{CR} {The value of the gap amplitudes obtained for this material scales relatively well with its $T_{c}$ in light of the recent results for the Fe-based superconductors.~\cite{Hafiez1,Yapono}} In addition, one can notice that the extracted ratio for anisotropic $s$-wave order parameter $\alpha$ is smaller than the BCS value, which points to existence of the large gap.

It is important to note that ARPES studies report two $s$-wave gaps of 2.3 and 7.8 meV for the outer and the inner Fermi surface sheets, respectively, without any nodes~\cite{DE}. In fact, ARPES results hint towards the conclusion about strong dependence of the gap value on orbital character of the bands forming the corresponding Fermi surfaces: the larger gap appears on d$_{xz}$/d$_{yz}$ bands.~\cite{AAK} Very recently, and based on a multi-band Eliashberg analysis, Ca$_{0.32}$Na$_{0.68}$Fe$_{2}$As$_{2}$ demonstrates that the superconducting electronic specific heat is well described by a three-band model with an unconventional $s_{\pm}$ pairing symmetry with gap magnitudes of approximately 2.35, 7.48, and -7.50 meV~\cite{STJ}. {It has been well demonstrated that the model based on Eliashberg equations is a simplified model of the real four bands model taking into account the similarities between the two 3D Fermi sheets and between the two 2D Fermi sheets. Based on them for the determination of $T_{c}$ and for the gap functions there can be considered only a distinct gap for every 2D, and respectively 3D sets of bands.~\cite{Dolgov} In fact, in order to solve the Eliashberg equations, there were two ways. The first one was to solve the equations which contain dependencies of real frequency, and the second one was to solve this equations on the imaginary axis, summing on Matsubara frequencies.~\cite{DJ} In contrast, the $\alpha$-model is not self-consistent, but provides a popular model with which experimentalists can fit their thermodynamic data that deviate from the BCS predictions and to quantify those deviations.~\cite{DC2}}

Although a clear picture is still missing for the case of Ca$_{0.32}$Na$_{0.68}$Fe$_{2}$As$_{2}$, it is important to emphasize that our system could be described via multiband superconductivity. However, from the temperature dependence of the lower critical field data alone it is difficult to be sure whether one, two or three bands can well describe our investigated system, since in the case of multiband superconductivity low-energy quasiparticle excitations can be always explained by the contribution from an electron group with a small gap.

\subsection{SNS-Andreev spectroscopy}

It is widely known that intrinsic multiple Andreev reflections effect (IMARE) develops on cryogenic clefts of some layered superconductors.~\cite{Pono} For example, such SnS-Andreev arrays were found in Gd-1111.~\cite{TEK3} IMARE spectroscopy is a powerful tool to determine bulk superconducting properties, that is why we use this method on Ba$_{0.65}$K$_{0.35}$Fe$_{2}$As$_{2}$ single crystals. The current-voltage characteristics, I(V), of the break junctions demonstrate features typical for SnS-Andreev mode. I(V) for one of these junctions, with excess current at low bias voltages (foot) is shown in Fig.\,5 by the black solid line. The foot area in I(V) is manifested in dI(V)/dV spectrum as a drastic increase of dynamic conductance. With it, the spectrum reveals series of peculiarities marked by n$_{L}$ labels and arrows, which should form subharmonic gap structure (SGS) described by Eq.\,1, and corresponds to the theories.~\cite{MOc,Flen,GBA,AKU} Although for high-transparent junctions the theory predicts the set of dynamic conductance minima, the peculiarity $\sharp$3 appears to be rather smeared, probably, due to the pronounced foot. Such nontrivial conductance raise may have multiband nature (several channels of Andreev transport in parallel) and is repeatedly observed on other Fe-based superconductor contacts.~\cite{TEK3,Yapono} To check whether the peculiarities form a SGS, we plot the dependence of their positions V$_{n}$ on inverse number 1/n (see lower inset of Fig.\,5). The linear dependence tending to the origin proves that the peculiarities belong to the same SGS. Theory for MARE by K\"{u}mmel {\it et al.},~\cite{AKU} suggests that the number of visible Andreev minima on dynamic conductance spectra is not less than the $l/a$ ratio. Consequently, for the contact from Fig.\,5 one could estimate $l \approx 2a$. The gap magnitude is hence determined as 2$\Delta$ = eV$_{n}$n $\approx$ 16.0 meV. Note that the presence of three peculiarities here increases the accuracy of the gap value obtained. The latter is close to the gap 2$\Delta \approx$ 20 meV measured in ARPES study of Ba$_{0.65}$K$_{0.35}$Fe$_{2}$As$_{2}$ single crystals from the same batch (see Fig.\,1c Ref.~\onlinecite{DVEv} for details).

\begin{figure}
\includegraphics[width=20pc,clip]{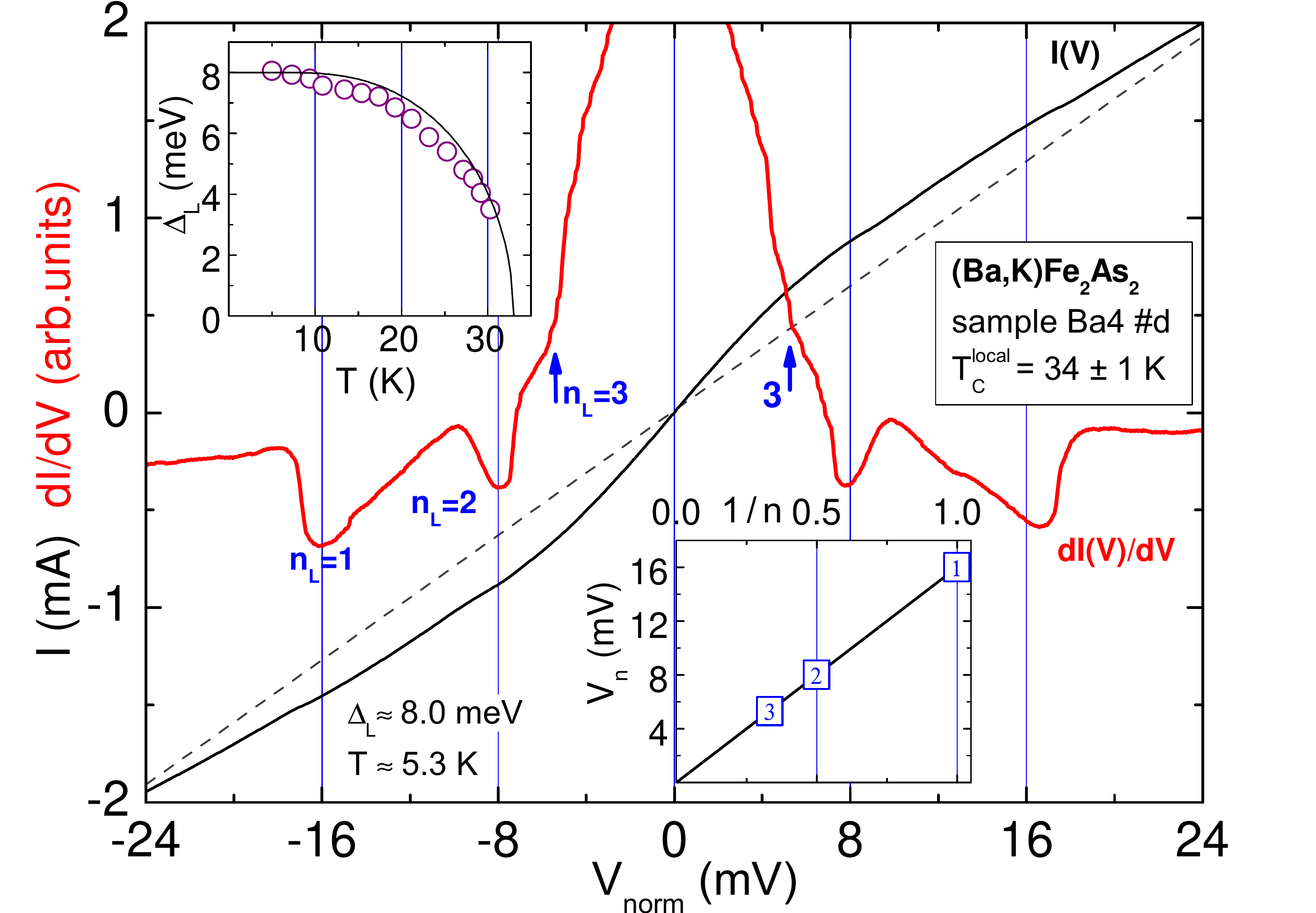}
\caption{Current-voltage characteristics I(V) and the dynamic conductance spectra dI(V)/dV for the SNS-Andreev array $\sharp d$ realized on Ba4 sample (two junctions in the stack). The bias voltage is scaled down by a factor of 2, correspondingly. The data were obtained at $T$ = 5.3 K. The local critical temperature of the contact point is about (34$\pm$1) K. Dashed line, crossing (0,0) point, is linear dependence plotted for comparison. Blue n$_{L}$ labels with arrows indicate the subharmonic gap structure (SGS). Upper inset shows experimental $\Delta_L(T)$ data (circles). The solid line is BCS-like dependence. Lower inset shows bias voltages V$_{n}$ for the n$_{L}$ series of dips versus their inverse  ordinary numbers. Note, that the V$_{n}$(1/n) dependence (line), as expected passes through the origin, evidencing for $\Delta_L\approx 8.0$\,meV.} \label{Fig:1}
\end{figure}

The temperature dependence of this gap shown in the upper inset of Fig.\,5 agrees well with BCS-like behavior (solid line) but slightly bends down. The latter is typical for nonzero interband interaction with another superconducting condensate described by a smaller gap $\Delta _{S}$ (see, for example,~\cite{SAK,TEK2}), which is to exist beyond the observed large gap $\Delta _{L}$. Therefore, the peculiarities observed are caused by the large gap $\Delta _{L}$. The small gap SGS located at lower biases seemed to be smeared by the foot. The dynamic conductance spectrum becomes flat at approximately 34\,K which corresponds to the termination of Andreev transport, thus defining the local critical temperature $T_{c}^{local}$ of the contact area. The latter allows us to calculate the BCS-ratio 2$\Delta _{L}$/k$_{B}$$T_{c}^{local} \approx$ 5.5 more exactly. Note that this value is the highest obtained in this work for Ba$_{0.65}$K$_{0.35}$Fe$_{2}$As$_{2}$.

Now we detail the shape of Andreev minima in Fig.\,5. Bearing in mind an exponential background, one could detect the reproducible shape of $n_L$=1 and $n_L$=2 peculiarities. Although the fine structure is smeared their slight asymmetry suggests that the $\Delta _{L}$ condensate is described by an extended $s$-symmetry. Since the SGS minima are rather pronounced, have substantial amplitude, and their line shape does not match the theoretically predicted one for $d$-wave case,~\cite{Flen,AP} we conclude on the absence of nodes in the $\Delta _{L}$ gap. A rough estimation gives about 30\% anisotropy in k-space. Similar degree of anisotropy could be attributed to gap peculiarities in the spectra presented in Fig.\,6.

\begin{table*}
\caption{\label{tab:table 1}  {The superconducting transition temperature $T_{c}$, and the superconducting gap properties extracted from IMARE and lower critical field ($H_{c1}$) studies for
Ba$_{0.65}$K$_{0.35}$Fe$_{2}$As$_{2}$ (three samples), and Ca$_{0.32}$Na$_{0.68}$Fe$_{2}$As$_{2}$, respectively along with other 122 Fe-based superconductors.}}
\begin{ruledtabular}
\begin{tabular}{cccccccc}
Compounds &$T_c$ (K) & Nodes, anisotropy & $\Delta _{L}(meV)$ & $\Delta _{S}(meV)$ & $\Delta _{L}/\Delta _{S}$ &  Technique &{Ref.}\\

\hline
Ba$_{0.65}$K$_{0.35}$Fe$_{2}$As$_{2}$$^{a}$\footnotetext[1]{sample Ba4, contact $\sharp$d (Ba4$\sharp$d)} &34$\pm$1 &no,$\approx$30\%  &8--4.8 & invisible &-- &break-junction &this work  \\
\hline
Ba$_{0.65}$K$_{0.35}$Fe$_{2}$As$_{2}$$^{b}$\footnotetext[2]{sample Ba4, contact $\sharp$b (Ba4$\sharp$b)} &34$\pm$3  &no,$\approx$25\%  &7.4--5.5 & invisible &-- &break-junction &this work  \\
\hline
Ba$_{0.65}$K$_{0.35}$Fe$_{2}$As$_{2}$$^{c}$\footnotetext[3]{sample Ba6, contact $\sharp$5 (Ba6$\sharp$5)}&34$\pm$3  &no,$\approx$25\%  & 7.4--5.7 & 1.7$\pm$0.3 &4.35$\pm$0.3 &break-junction &this work  \\
\hline
Ca$_{0.32}$Na$_{0.68}$Fe$_{2}$As$_{2}$ &34$\pm$1&no  &7.28$\pm$0.3 & 2.26$\pm$0.3&3.22$\pm$0.3 &magnetization & this work  \\
\hline
Ba$_{0.6}$K$_{0.4}$Fe$_{2}$As$_{2}$ &35.8 &no  &8.9$\pm$0.4 & 2.0$\pm$0.3 &4.45$\pm$0.3 &magnetization&~\cite{CR}\\

\end{tabular}
\end{ruledtabular}
\end{table*}

The I(V) curves in Fig.\,6 are rather straight and have less pronounced Andreev peculiarities than those in Fig.\,5. It is generally supposed that such the suppression of Andreev excess current happens because of inelastic scattering process at the NS interfaces. In our case of $l/a > 1.5$ and atomically flat cryogenic clefts the inelastic scattering should be not a crucial reason for I(V) flattening. The plausible cause may be the presence of accidental atoms on the cryogenic clefts, which decreases the contact transparency. If accidental atoms are magnetic (Fe), such the magnetic centers could polarize electron spins and in this way can prevent the electron to find the pair during Andreev reflection process. Nevertheless, our experimental setup and the current modulation method we used are sensitive enough to obtain the details of dynamic conductance spectra. The main SGS minima n$_{L}$ = 1 being rather wide, form doublets caused by anisotropy of about 25\%. The upper curve in Fig.\,6 is for the contact obtained in Ba6 sample and the lower black curve corresponds to Ba4 sample from the same batch. A smooth background is suppressed here for both spectra. Easy to see that the fine structure of n$_{L}$ = 1 minima is well-reproduced, whereas the CVC's presented (thin lines) show a considerably different contact resistances, and, therefore, the contact areas.

\begin{figure}
\includegraphics[width=20pc,clip]{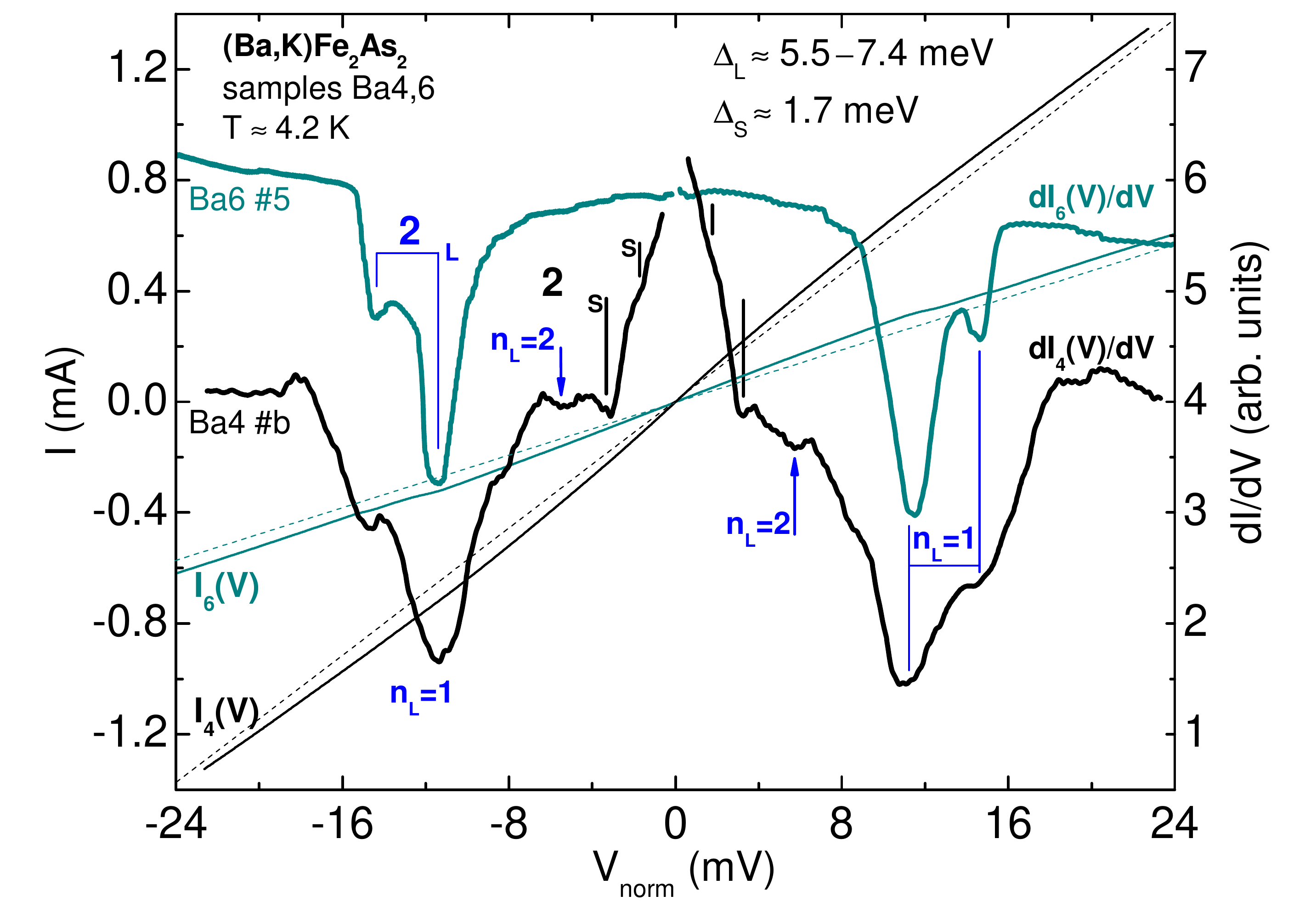}
\caption{Current-voltage characteristics I(V) (thin) and their dynamic conductance spectra dI(V)/dV (bold lines) for two SNS-Andreev arrays $\sharp5$ and $\sharp b$ realized in Ba6 and Ba4 samples from the same batch. Since there are two junctions in the stack, the bias voltage is scaled down by a factor of 2, $T$ = 4.2\,K. The local critical temperature of the contact point is about (34$\pm$3) K. Dashed linear dependencies, crossing the origin, are plotted for the comparison. Blue labels and arrows indicate the subharmonic gap structure (SGS) of $\Delta _{L}$, black labels and vertical dashes -- SGS of $\Delta _{S} \approx$ (1.7$\pm$0.3)\,meV.  $\Delta _{L} \approx$  (5.5-7.4) meV, the width of Andreev minima is determined by the k-space anisotropy about 30\%, and signed by the marker lines. Monotonic background of conductance spectra is subtracted.} \label{Fig:1}
\end{figure}

The spectra shown in Fig.\,6 demonstrate the average value of the large gap $\Delta _{L} \approx$ 6.5 meV, and the resulting BCS-ratio 2$\Delta _{L}$/k$_{B}$$T_{c}^{local} \approx$ 4.5. The anisotropy causes the  $\Delta _{L}$ smearing in the range from 5.5 to 7.4 meV. The lower threshold is in a good agreement with the minimal value of the hole-band gap from.~\cite{DVEv} The lower spectrum labeled as Ba4 $\sharp$b has more intensive peculiarities at low bias. Here, the second Andreev minima n$_{L}$ = 2 are resolved (marked by arrows). The next minima have higher amplitude, thus beginning the new SGS. Therefore, the spectrum contains Andreev structure set by the small gap  $\Delta _{S} \approx$ 1.7 meV (marked by vertical dashes). Obviously, additional studies are needed for more accurate determination of the small gap.

In the majority of the spectra obtained for Ba$_{0.65}$K$_{0.35}$Fe$_{2}$As$_{2}$ we also observed less pronounced peculiarities at bias voltages V $>$ 2$\Delta _{L}$/e (not shown here). It is interesting to note that the shape of the lowest spectra in Fig.\,6 for contact $\sharp$b is similar to that of LiFeAs contact $\sharp$d2 obtained by us earlier (see Fig.\,1, dashed curve in Ref.~\onlinecite{TEK2} for comparison). Both facts point to a possible presence of a third, the largest superconducting gap developing in the bands with a vanishing density of states. As for the large gap observed, the BCS-ratio 2$\Delta _{L}$/k$_{B}$$T_{c} \approx$ 5$\pm$0.5 indicates a strong electron-boson pairing in $\Delta _{L}$ band.

For the sake of comparison, we have summarized the values of gaps $\Delta _{L}$, $\Delta _{S}$, and $T _{c}$ for  Ba$_{0.65}$K$_{0.35}$Fe$_{2}$As$_{2}$ extracted from IMARE and for Ca$_{0.32}$Na$_{0.68}$Fe$_{2}$As$_{2}$ extracted from magnetization measurements along with other hole-doped 122 materials in Table I. For both investigated systems, the large gap $\Delta_L$ has a higher value than the weak-coupling BCS (1.76$k_{B}T_{c}$) gap value, which reflects a tendency for strong coupling effects, while the smaller one $\Delta_S$ has a value lower than the BCS one. Our Table points that $\Delta _{L}$/$\Delta _{S}$ ratio is nearly constant. The mentioned ratio of the investigated systems in this paper is not surprising being comparable with optimally doped Ba$_{0.6}$K$_{0.4}$Fe$_{2}$As$_{2}$,~\cite{CR}  and Ca$_{0.32}$Na$_{0.68}$Fe$_{2}$As$_{2}$~\cite{STJ} but disagrees with earlier SH measurements for lower-$T_{c}$ Ba$_{0.65}$Na$_{0.35}$Fe$_{2}$As$_{2}$ sample;~\cite{AK} possible reasons for this inconsistency were mentioned above. Although, the gap values are scattered for different compounds within the 122 family, our obtained gap structure has qualitative similarity and comparable with the two-band $s$-wave fit for the lower critical field data of Ba$_{0.6}$K$_{0.4}$Fe$_{2}$As$_{2}$.~\cite{CR}

A puzzling issue that our results rise is the potential presence of the 3$^{rd}$ gap in the 122 systems. To address  this issue and to have a deeper insight on the gap anisotropy,  high-precision ARPES, or low-temperature STM data would be highly desirable though they are currently challenging.



\subsection{Conclusions}

Using complementary experimental techniques, we studied  single crystals of  the 122-FeAs family, and obtained consistent data on their superconducting order parameter. From the previous detailed analysis, the temperature dependence of the $\lambda^{-2}_{ab}(T)$ is inconsistent with a simple isotropic $s$-wave type of the order parameter but are rather in favor of the presence of a two $s$-wave-like gaps or an anisotropic s-wave. These observations clearly show that the superconducting energy gap in Ca$_{0.32}$Na$_{0.68}$Fe$_{2}$As$_{2}$ is nodeless. In addition, the gaps obtained from our $H_{c1}$ measurements are clearly similar to those determined from the ARPES measurements. The IMARE spectroscopy of SNS-Andreev arrays formed by the break-junction technique reveals two nodeless gaps: the large gap,  $\Delta _{L}$ = (6--8) meV (depending on small variation of $T_{c} \approx$ 34 K) with extended $s$-wave symmetry, anisotropy in the $k$-space not less than $\approx$ 30\%, and the small gap, $\Delta _{S}$ = 1.7$\pm$0.3 meV. According to our SNS-Andreev data, the BCS-ratio for the large gap is 2$\Delta _{L}$/k$_{B}T_{c} \approx$  5$\pm$0.5.


\begin{acknowledgments}
The authors thank Roman Kramer, Anja Wolter-Giraud, Shigeto Hirai, Yaroslav G. Ponomarev, Joris Van de Vondel, Bernd B\"{u}chner and Sabine Wurmehl  for fruitful discussions. This work is supported by the FNRS projects, ``cr\'edit de d\'emarrage U.Lg.", the MP1201 COST Action and by the Methusalem Funding of the Flemish Government. By the Russian Foundation for Basic Research (project nos. 13-02-01180-a, 14-02-90425, 14-02-92002-a), and Civilian Research Development Foundation grant FSAX-14-60108-0 as well as Russian Ministry of Education and Science, Russian Science Foundation, and NAS of Ukraine (project 73-02-14) for support.
\end{acknowledgments}

\end{document}